\documentclass[journal]{IEEEtran}

\usepackage{xcolor,soul,framed}
\colorlet{shadecolor}{yellow}
\usepackage[pdftex]{graphicx}
\DeclareGraphicsExtensions{.pdf,.jpeg,.png}
\usepackage{amsfonts,amssymb}
\usepackage{mathrsfs}
\usepackage{amsthm}
\usepackage[cmex10]{amsmath}
\usepackage{mathtools}
\usepackage{bbm}
\DeclareMathAlphabet{\mathbbb}{U}{bbold}{m}{n}
\usepackage{array}
\usepackage{flushend}
\usepackage{physics}
\usepackage{cite}
\usepackage{comment}
\usepackage{tikz}
\usetikzlibrary{arrows.meta,positioning,calc}
\usepackage{booktabs}
\usepackage{multirow}
\usepackage{tabularx}
\usepackage[caption=false,font=footnotesize]{subfig}
\usepackage{algorithm}
\usepackage{algpseudocode}
\usepackage{siunitx}
\usepackage{stfloats}
\usepackage{balance}
\usepackage{hyperref}
\hypersetup{hidelinks}

\theoremstyle{remark}

\algrenewcommand\algorithmicrequire{\textbf{Input:}}
\algrenewcommand\algorithmicensure{\textbf{Output:}}

\newcommand{\tabincell}[2]{\begin{tabular}{@{}#1@{}}#2\end{tabular}}

\newcommand{\trans}{{}^{\mathsf{T}}}
\newcommand{\pha}[1]{\underline{#1}}

\begin{document}

\title{Seamless Contraction-Control Framework for Unplanned Grid-Connected/Stand-Alone Transitions of Grid-Forming Inverters}

\author{Qianxi~Tang,~\IEEEmembership{Student Member,~IEEE,}
	and~Li~Peng,~\IEEEmembership{Senior Member,~IEEE}%
	\thanks{This paper has not been presented elsewhere previously. The authors are with School of Electrical and Electronic Engineering, Huazhong University of Science and Technology, Wuhan 430074, China. Email: qianxi@hust.edu.cn, pe105@mail.hust.edu.cn}}

\maketitle

\begin{abstract}
Unplanned grid-connected (GC)/stand-alone (SA) transitions commonly occur in AC microgrids during protection trips, manual breaker operation, or low-bandwidth supervisory communication. Under such unplanned transitions, a grid-forming inverter must support the local-load voltage in stand-alone operation and regulate the desired power/current injection in grid-connected operation. Existing P--Q droop-based seamless-transfer methods often rely on planned transition commands, supervisory islanding detection, or pre-synchronization interval, which may prevent timely voltage/current support during unplanned bidirectional transitions. To address this problem, this paper proposes a seamless contraction-control (SCC) framework for target dynamics. Using the SCC, contraction-based grid-connected current-control and stand-alone voltage-control laws are proposed. With the new control laws, the inverter achieves transient stability and converges to the target trajectory with a prescribed convergence rate. Furthermore, a breaker-status observer is proposed to infer the grid-connected/stand-alone mode from voltage measurements on both sides of the breaker, eliminating the need for a dedicated pre-synchronization interval or supervisory islanding detection process and enabling timely voltage/current support during unplanned transitions. Experimental results validate that the proposed method achieves stand-alone voltage support, stable grid-connected current injection under symmetrical/unsymmetrical grid-voltage sag and phase-jump disturbances, and unplanned bidirectional transitions.
\end{abstract}


\begingroup
\small
\setlength{\parskip}{0pt}
\setlength{\parindent}{0pt}
\newlength{\nomleft}
\newlength{\nomright}
\setlength{\nomleft}{2.35cm}
\setlength{\nomright}{\dimexpr\columnwidth-\nomleft-0.18cm\relax}
\newcommand{\nomentry}[2]{%
	\noindent\parbox[t]{\nomleft}{\raggedright #1\strut}%
	\hspace{0.12cm}%
	\parbox[t]{\nomright}{\raggedright #2\strut}%
	\par\vspace{0.25ex}}
\noindent\makebox[\columnwidth][c]{\scshape Nomenclature}\par

\nomentry{$\pha{(\cdot)}$}{Three-phase instantaneous vector in the $ABC$ frame}
\nomentry{$(\cdot)^\star$}{Reference command}
\nomentry{$\widehat{(\cdot)}$}{Estimated quantity}
\nomentry{$(\cdot)^{\trans}$}{Matrix or vector transpose}
\nomentry{$k$}{Discrete-time sample index}
\nomentry{$\pha{u}$}{Full-bridge switch output voltage vector}
\nomentry{$\pha{i}_1$}{Inverter-side inductor current vector}
\nomentry{$\pha{v}_c$}{Filter-capacitor voltage vector}
\nomentry{$\pha{i}_2$}{Grid-side or inverter output current vector}
\nomentry{$\pha{v}_{\mathrm{PCC}},\pha{v}_b$}{PCC-side and inverter-side breaker-voltage vectors}
\nomentry{$L_1,L_2,C_f$}{Plant filter inductances and capacitance}
\nomentry{$p,q$}{Instantaneous active and reactive powers}
\nomentry{$P,Q,S$}{Filtered active power, filtered reactive power, and apparent power}
\nomentry{$P_{\mathrm{ref}},Q_{\mathrm{ref}}$}{Active- and reactive-power references}
\nomentry{$V_r,\omega_r,\theta_r$}{Droop voltage, frequency, and phase references}
\nomentry{$K_P,K_Q$}{M1 $P$--$f$ and $Q$--$V$ droop gains}
\nomentry{$K_{p,u},K_{i,u}$}{Voltage-loop PI gains used in M1 and M2}
\nomentry{$\tau_p$}{Power-filter time constant}
\nomentry{$\chi_g$}{Grid-connected mode flag used in M2}
\nomentry{$\psi_Q$}{Reactive-power integral state in the M2 droop loop}
\nomentry{$n,m,m_{\mathrm{int}}$}{M2 active, reactive, and integral droop gains}
\nomentry{$\pha{x}_m,\pha{r}_m$}{Regulated variable and reference in mode $m$}
\nomentry{$\pha{e}_m,\pha{s}_m,\bar{\pha{s}}_m$}{Tracking error and target-dynamics auxiliary variables}
\nomentry{$\pha{f}_m,\pha{g}_m$}{Feedback-shaped plant dynamics and target dynamics}
\nomentry{$\Gamma_m,k_{m,j}$}{Auxiliary contraction gain matrix and target-polynomial coefficients}
\nomentry{$e,x,r$}{Representative scalar tracking error, state, and reference}
\nomentry{$\pha{g}_v,\pha{g}_i$}{Stand-alone-voltage and grid-connected-current target signals}
\nomentry{$\pha{u}_{\mathrm{off}},\pha{u}_{\mathrm{on}}$}{Stand-alone and grid-connected control inputs}
\nomentry{$D_{\tau}(s),\tau_d$}{Filtered differentiator and derivative-filter time constant}
\nomentry{$k_{v1},k_{v0}$}{Stand-alone voltage-control error gains}
\nomentry{$\omega_n,\zeta$}{Stand-alone voltage-law natural frequency and damping ratio}
\nomentry{$k_{i2},k_{i1},k_{i0}$}{Grid-connected current-control error gains}
\nomentry{$\lambda_{1,2,3}$}{Grid-connected current-law pole rates}
\nomentry{$\pha{\sigma}$}{Estimated Status from the proposed breaker-status observer}
\nomentry{$e_A,e_B,e_C$}{Instantaneous breaker-voltage mismatch}
\nomentry{$E_{\mathrm{th}}$}{Breaker-status-observer voltage threshold}
\nomentry{$N_w,N_c,c_{\phi}$}{Moving-average length, persistence count, and counter}
\par\endgroup

\section{Introduction}
\IEEEPARstart{W}{ith} the increasing penetration of renewable energy resources, distributed generators, and power-electronic interfaced equipment, AC microgrids have become an important form of modern power systems. In application scenarios such as isolated power supply, distributed generation integration, and resilient local energy support, grid-forming inverters (GFM) can actively establish voltage and frequency and therefore have broad application prospects \cite{tian2023transient_sync,wang2023inertia_emulation}. Under a fixed electrical topology, this voltage-source behavior provides a basis for both local-load supply and grid-supporting operation. However, practical GFM inverters experience both planned and unplanned transitions between grid-connected and stand-alone operation. During such transitions, abrupt changes in network structure and control objectives may induce severe voltage fluctuation, current surge, and oscillatory transients, which can threaten power quality and equipment safety \cite{ferrari2024grid_forming,sun2025enhanced_transient}.

\begin{figure}[!t]
	\centering
	\includegraphics[width=0.98\columnwidth]{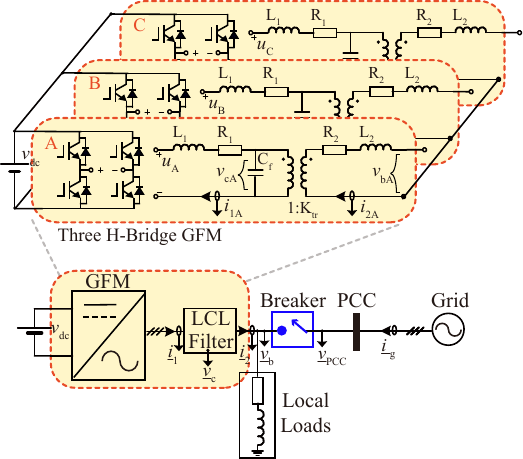}
	\caption{Studied complete system of a grid-forming converter with local measurements for breaker-status observation.}
	\label{fig:topology}
\end{figure}

Existing seamless-transfer studies can be grouped according to how they obtain mode information and how they prepare the inverter for the next operating state. One widely used class relies on islanding/reconnection detection, transfer commands, or pre-synchronization \cite{vandoorn2013transition_droop,amin2020resynchronization_udc,ref7,ref10,song2026series_flexible}. In the pre-synchronized subset, the stand-alone inverter or microgrid voltage is adjusted before reconnection so that the voltage magnitude, frequency, phase sequence, and phase angle satisfy a closing window. Such a process is effective for planned reconnection, but it normally requires a supervisory decision, breaker-side status information, grid-voltage phase information, or a communication path among the breaker, synchronizing unit, and inverters \cite{arafat2015smooth_transition,wang2015generalized_transition,wang2019integrated_sync,wang2020nonpll_transition,westman2021goose_transition,vukojevic2020microgrid_protection,ref10}. Therefore, its benefit decreases when the breaker action is caused by protection trips, external reclosing, manual operation, delayed auxiliary contacts, or other events in which the electrical topology changes before a validated synchronization command is available. A second class uses dual-mode or droop-based controllers to smooth the transition between grid-connected and stand-alone operation \cite{hou2018distributed_hierarchical,buduma2022seamless_master_slave,ref8,sadeque2024autonomous}. These methods improve operating-mode adaptability, but the grid-connected power target is usually reached through voltage magnitude, frequency, phase, or droop dynamics. As a result, the post-transition power injection requires droop-loop settling and depends on the network impedance and post-event voltage condition after reconnection. This limitation is especially important when the inverter must not only preserve the local-load voltage after breaker opening, but also rapidly inject a commanded grid-service current after reconnection. A third group reduces communication dependence using local voltage/frequency quantities or current-limiting actions \cite{ref8,ref9,ref10}. Local variables are attractive for practical inverters, and current-limiting or protection-oriented methods can keep current within hardware limits during severe voltage disturbances or reconnection events \cite{huang2019transient_current_limit,qoria2020current_limiting,ferrari2024grid_forming,sun2025enhanced_transient}. Recent no-detector or no-pre-synchronization studies further improve autonomous operation, including phase-adaptive current limiting for stand-alone-to-grid-connected reconnection \cite{yalaoui2024seamless,guo2026seamless_gfm_pv}. However, stand-alone-takeover methods mainly address one direction, while current limiting is a protective constraint and does not assign the desired voltage/current recovery trajectory. The remaining issue is a local method that identifies the post-event mode and assigns recovery dynamics for arbitrary bidirectional grid-connected/stand-alone transitions.

 ``Breaker-status communication is unavailable'' means that the inverter does not receive a validated breaker command, islanding decision, or synchronizing phase reference before the breaker changes the electrical topology. This does not exclude local voltage/current measurements. It covers low-bandwidth islanding decisions \cite{ref7,ref9}, delayed auxiliary contacts, protection trips, external reclosing, and manual breaker operations, which define the unplanned-transition problem considered in this paper. This problem is difficult because an unplanned seamless grid-connected/stand-alone transition is not only a steady-state power-flow change. First, the $LCL$-filtered inverter is a high-order plant, and stored filter energy can cause oscillatory current and capacitor-voltage transients if the internal states are not directly shaped \cite{pan2017current_control_lcl,wang2017harmonic_instability,hu2023generic}. Second, delayed breaker or islanding information makes the event an unknown boundary-condition and robust-control problem at the transition instant \cite{ref7,ref9}. Third, breaker actions and large grid-voltage disturbances, such as voltage sag and phase jump, impose large-signal transient-stability requirements because pre-event references may become inconsistent with the post-event electrical constraint \cite{huang2019transient_current_limit,qoria2020current_limiting,ferrari2024grid_forming,sun2025enhanced_transient}.

Motivated by these issues, this paper proposes a seamless contraction control (SCC) framework for unplanned bidirectional grid-connected/stand-alone transitions of grid-forming inverters and a breaker-status observer for local transition-mode identification. Based on the SCC, two new controllers are derived: a grid-connected current controller and a stand-alone voltage-forming controller. The SCC framework and observer are demonstrated for the complete unplanned bidirectional grid-connected/stand-alone operating process. The main contributions of this work are:
\begin{enumerate}
\renewcommand{\labelenumi}{\arabic{enumi})}
	\item The SCC framework for target dynamics is to maintain stable voltage/current behavior of an $LCL$-filtered grid-forming inverter under large-signal disturbances, including symmetrical/unsymmetrical grid-voltage sag, phase jump, and unplanned bidirectional grid-connected/stand-alone transitions.
	\item The SCC assigns desired high-order $LCL$-filter dynamics for fast voltage/current recovery with expected convergence rates under the above large-signal disturbances. Controller-tuning methods are also provided.
	\item The breaker-status observer is integrated with the SCC controllers to realize arbitrary unplanned bidirectional grid-connected/stand-alone transitions, without preparation before reconnection, a pre-synchronization interval, supervisory islanding detection, or breaker-status communication.
\end{enumerate}
\begin{comment}
The remainder of this paper is organized as follows. Section II introduces the system configuration, averaged model, and mode-transition problem. Section III presents the SCC controller, target-dynamics tuning, and breaker-status observer. Section IV reports the hardware experiment and results. Section V concludes the paper.
\end{comment}
\section{System Configuration and Problem Formulation}
\subsection{GFM system configuration for mode transitions}
\begin{figure*}[!t]
	\centering
	\includegraphics[width=0.9\textwidth]{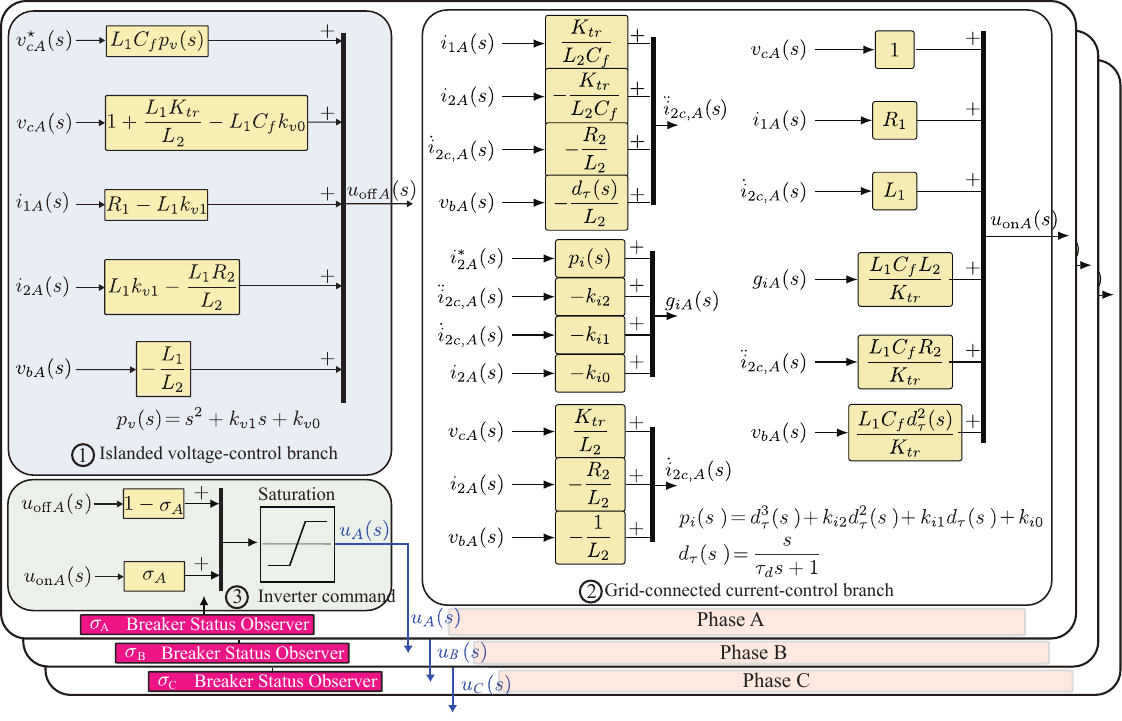}
	\caption{Proposed seamless contraction control framework for unplanned grid-connected/stand-alone transitions: grid-connected current-control law and stand-alone voltage-control law.}
	\label{fig:control_block}
\end{figure*}

Fig.~\ref{fig:topology} shows the three-phase electrical structure in the $ABC$ frame and the measured variables used throughout this paper. The studied grid-forming inverter plant consists of a three-phase full-bridge converter supplied by a fixed dc source, an $L_1C_fL_2$ filter, and a transformer. The inverter is connected to the utility grid through a breaker at the PCC, and the voltages on both sides of the breaker are measured and fed back to the controller for breaker-status observation. The notation used in the proposed SCC controller is shown in Fig.~\ref{fig:control_block}. Underlined symbols denote three-phase instantaneous vectors in the $ABC$ frame; for example, $\pha{i}_1=[i_{1A},i_{1B},i_{1C}]^\mathrm{T}$ and $\pha{v}_c=[v_{cA},v_{cB},v_{cC}]^\mathrm{T}$. Scalar phase components, such as $v_{cA}$, and scalar parameters, such as $L_1$ and $R_1$, are written without underlines. Accordingly, the inverter-side inductor current, filter-capacitor voltage, grid-side current, PCC-side breaker-voltage vector, and estimated breaker status are denoted by $\pha{i}_1$, $\pha{v}_c$, $\pha{i}_2$, $\pha{v}_{\mathrm{PCC}}$, and $\pha\sigma$, respectively.

\subsection{Unplanned Bidirectional Transition Problem}
 The averaged vector model of the studied LCL GFM inverter can be written as
\begin{align}
L_1 \dot{\pha{i}}_1 &= \pha{u} - R_1 \pha{i}_1 - \pha{v}_c,
\label{eq:model_i1}\\
C_f \dot{\pha{v}}_c &= \pha{i}_1 - \pha{i}_2,
\label{eq:model_vc}\\
L_2 \dot{\pha{i}}_2 &= K_{tr}\pha{v}_c - R_2 \pha{i}_2 - \pha{v}_b,
\label{eq:model_i2}
\end{align}
where $\pha{u}$ is the full-bridge switch output voltage vector or the control input and $\pha{v}_b$ represents the inverter-side breaker-voltage vector, i.e., the boundary voltage applied at the $L_2$ terminal by the breaker status, grid condition, and local network. Fig.~\ref{fig:topology} shows where these variables are measured and how they are defined.

Equations \eqref{eq:model_i1}--\eqref{eq:model_i2} show why the transition is not a simple reference switch. In stand-alone operation, $\pha{v}_c$ is regulated and $\pha{i}_2$ acts as the load-current disturbance to \eqref{eq:model_vc}, possibly with load-current feedforward or output-impedance compensation \cite{lin2022load_current_feedforward,lin2023harmonic}. In grid-connected operation, $\pha{i}_2$ is regulated and $\pha{v}_b$ is the grid-imposed boundary voltage in \eqref{eq:model_i2}. Therefore, a grid-connected-to-stand-alone event removes the stiff grid boundary and changes a current/power objective into a local-voltage support objective, whereas a stand-alone-to-grid-connected event imposes the utility-grid voltage and requires rapid recovery of the desired grid-service current. During an unplanned breaker action, this objective change and the model boundary-condition change can occur before a validated status signal is available.

These equations expose the coupled control difficulty during the unplanned bidirectional transition interval. The $LCL$ filter is a high-order energy-storage network, so the inductor-current and capacitor-voltage states can carry oscillatory energy across the breaker event and must be shaped directly to satisfy damping and settling requirements. At the same time, the breaker-side boundary in $\pha{v}_b$ may switch between a grid-imposed voltage and a local-network condition before validated breaker or islanding information reaches the controller, which makes the transition an unknown boundary-condition and robust-control problem. When the breaker action is accompanied by large grid-voltage disturbances, such as voltage sag or phase jump, the same current command corresponds to a different instantaneous switch-bridge voltage output; therefore, the post-event voltage and current trajectories make the transition a transient-stability-related problem, rather than a pre-event power-matching problem alone. Otherwise, pre-event current/voltage references may no longer satisfy the post-event electrical constraint and can produce PCC-voltage fluctuation or current surge.

\subsection{Baseline Communicated Transition and Its Limitation}
The traditional P-Q droop communicated transition process is formulated as a communicated transition problem. As illustrated in Fig.~\ref{fig:baseline_droop_process}, a supervisory island-detection unit first decides whether the microgrid is grid-connected or stand-alone and, in some implementations, provides only the permission logic for opening or closing the breaker. If pre-synchronization is included, a separate unit actively aligns the stand-alone inverter or microgrid voltage with the utility grid before closing. The grid-side voltage magnitude, frequency, phase rotation, and phase angle are communicated to the inverter or to this pre-synchronization unit. The inverter then adjusts its stand-alone voltage until the utility-grid voltage and microgrid voltage satisfy the reconnection window; only after this preparation can the breaker be closed and the P-Q droop loop provide grid-supporting active/reactive power. Following IEEE Std. 1547.4-2011, this reconnection check includes acceptable voltage, frequency, phase rotation, and phase-angle-difference conditions, while the conditions depend on the applicable grid code and the nominal system frequency \cite{ieee1547_4_2011}. In contrast, the proposed SCC uses local voltage measurements on the two breaker sides to infer the operating state and select the corresponding contraction-derived control law without a communicated breaker-status signal.

The droop baselines in Fig.~\ref{fig:baseline_droop_process} are denoted as \textbf{M1} and \textbf{M2}. M1 is the traditional communicated P-Q droop controller \cite{vasquez2009adaptive_droop,deng2017enhanced_power_flow}; M2 is the mode-dependent droop controller in \cite{ref7}. For both methods, the instantaneous powers are computed as $p=\frac{3}{2}(v_{od}i_{od}+v_{oq}i_{oq})$ and $q=\frac{3}{2}(v_{oq}i_{od}-v_{od}i_{oq})$, and the filtered powers are obtained from $\tau_p\dot P=p-P$ and $\tau_p\dot Q=q-Q$. Other related variables are summarized in the Nomenclature. For M1, the conventional P-f and Q-V droop laws map active- and reactive-power errors directly to frequency and voltage references:
\begin{equation}
\begin{aligned}
\omega_r-\omega_{\mathrm{nom}} &=
K_P\!\left(P_{\mathrm{ref}}-P\right),\\
V_r-V_{\mathrm{nom}} &=
K_Q\!\left(Q_{\mathrm{ref}}-Q\right),\qquad
\dot\theta_r=\omega_r.
\end{aligned}
\label{eq:m1_traditional_pq_droop}
\end{equation}
For M2, $\chi_g$ is a constant tuning parameter that determines how the power reference is applied in the droop law. The P--f and Q--V laws are
\begin{equation}
\begin{aligned}
\omega_r &=
\omega_{\mathrm{nom}}
-n\!\left[P-(1-\chi_g)P_{\mathrm{ref}}\right],\\
V_r &=
V_{\mathrm{nom}}
-m\!\left[Q-(1-\chi_g)Q_{\mathrm{ref}}\right]
-\chi_g m_{\mathrm{int}}\psi_Q,\\
\dot\psi_Q &= Q-(1-\chi_g)Q_{\mathrm{ref}},\qquad
\dot\theta_r=\omega_r.
\end{aligned}
\label{eq:m2_mode_dependent_droop}
\end{equation}
M1 obtains the grid-service command through communicated references and the communicated transition path in Fig.~\ref{fig:baseline_droop_process}; M2 adds mode-dependent reference selection through $\chi_g$. Both methods regulate power through droop dynamics, so additional settling time is required before the desired post-transition power support is established.
\section{Seamless Contraction Control for Unplanned Mode Transitions}
\subsection{Seamless contraction-control framework for target dynamics}
Fig.~\ref{fig:control_block} summarizes the implemented control architecture. Following the full-order target-dynamics viewpoint in contraction control \cite{tsukamoto2021contraction}, SCC derives two mode-dependent control laws by assigning contracting post-event dynamics to the controlled variable of the active mode. In stand-alone operation, the stand-alone voltage-control law regulates the filter-capacitor voltage, with
\begin{equation}
	\pha{x}_{\mathrm{off}} = \pha{v}_c, \qquad
	\pha{r}_{\mathrm{off}} = \pha{v}_c^{\star}.
	\label{eq}
\end{equation}
The  PCC-voltage command could be the reference.
$\pha{v}_c^{\star}=\pha{v}_{\mathrm{PCC}}$.
In grid-connected operation, the grid-connected current-control law regulates the grid-side current, with
\begin{equation}
	\pha{x}_{\mathrm{on}} = \pha{i}_2, \qquad
	\pha{r}_{\mathrm{on}} = \pha{i}_2^{\star}.
	\label{eq}
\end{equation}
For each active mode $m\in\{\mathrm{off},\mathrm{on}\}$, the feedback-shaped plant is viewed as an $n_m$th-order system,
\begin{equation}
 \pha{x}_m^{(n_m)}
 =\pha{f}_m\!\left(\pha{x}_m,\dot{\pha{x}}_m,\ldots,\pha{x}_m^{(n_m-1)},\pha{u},t\right),
 \label{eq:actual_dynamics}
\end{equation}
and the controller assigns a target contracting dynamics of the same order,
\begin{equation}
 \pha{x}_m^{(n_m)}
 =\pha{g}_m\!\left(\pha{x}_m,\dot{\pha{x}}_m,\ldots,\pha{x}_m^{(n_m-1)},\pha{r}_m,t\right).
 \label{eq:target_dynamics}
\end{equation}
As in the full-order target-dynamics construction, define the auxiliary variable
\begin{equation}
 \pha{s}_m=\pha{x}_m^{(n_m-1)}-\bar{\pha{s}}_m,
 \label{eq:composite_s}
\end{equation}
with $\dot{\bar{\pha{s}}}_m=-\Gamma_m\!\left(\bar{\pha{s}}_m-\pha{x}_m^{(n_m-1)}\right)+\pha{g}_m$ and $\Gamma_m=\Gamma_m^{\trans}>0$, where $\Gamma_m$ is the positive-definite auxiliary contraction gain matrix for the $\pha{s}_m$ dynamics. Then $\pha{x}_m^{(n_m)}-\pha{g}_m=\dot{\pha{s}}_m+\Gamma_m\pha{s}_m$. This suggests that the nominal uncertainty-free dynamics of $\pha{s}_m$ should be selected as $\dot{\pha{s}}_m=-\Gamma_m\pha{s}_m$, although other contracting choices are possible provided that the nominal residual $\dot{\pha{s}}_m+\Gamma_m\pha{s}_m$ tends to zero. In this paper, $\pha{g}_m$ is selected as the linear tracking target
\begin{equation}
 \pha{g}_m=\pha{r}_m^{(n_m)} - \sum_{j=0}^{n_m-1} k_{m,j}\bigl(\pha{x}_m^{(j)}-\pha{r}_m^{(j)}\bigr),
\label{eq:general_target}
\end{equation}
where the coefficients are chosen such that the polynomial
$p_m(s)=s^{n_m}+\sum_{j=0}^{n_m-1}k_{m,j}s^j$ is Hurwitz. Let
$\pha{e}_m=\pha{x}_m-\pha{r}_m$. Substituting \eqref{eq:general_target}
into \eqref{eq:target_dynamics} gives the target error dynamics
\begin{equation}
	\pha{e}_m^{(n_m)}+\sum_{j=0}^{n_m-1}k_{m,j}\pha{e}_m^{(j)}=0 .
	\label{eq:target_error_dynamics}
\end{equation}
According to the Routh stability criterion \cite{routh1877stability}, the
Hurwitz condition on $p_m(s)$ makes the origin of \eqref{eq:target_error_dynamics}
exponentially stable. Therefore, $\pha{e}_m(t)\rightarrow0$, i.e.,
$\pha{x}_m(t)\rightarrow\pha{r}_m(t)$, and the linear target dynamics are
contracting in a constant metric \cite{tsukamoto2021contraction}. When the controller realizes
$\dot{\pha{s}}_m=-\Gamma_m\pha{s}_m$, the
feedback-shaped plant follows the target dynamics $\pha{g}_m$; hence the
closed-loop trajectory is governed by the assigned contracting error dynamics.

The linear auxiliary dynamics for $\pha{s}_m$ are not the only admissible choice. A more general contracting option is $\dot{\pha{s}}_m=\pha{\phi}_m(\pha{s}_m,t)$, where $\pha{\phi}_m$ is a smooth auxiliary vector field whose symmetric Jacobian with respect to $\pha{s}_m$ is uniformly negative definite. Other choices, such as nonlinear state-dependent damping or filtered target dynamics, can also be selected when a different transient response is required. This flexibility is useful in seamless-transfer operation because the preferred post-event response is not always the fastest convergence. Under weak-grid, high-load, or nonlinear-load conditions, the designer may choose a softer transient, bounded current slew rate, or stronger disturbance filtering while retaining contraction of the selected target dynamics.

\subsection{A new stand-alone voltage-control law}
The first contraction-derived SCC law is the stand-alone voltage-control law. It assigns the second-order target dynamics of $\pha{v}_c$ and uses the smooth voltage reference triplet $(\pha{v}_c^{\star},\dot{\pha{v}}_c^{\star},\ddot{\pha{v}}_c^{\star})$. The required plant terms are evaluated from measured states as
\begin{align}
 \dot{\pha{v}}_c &= \frac{\pha{i}_1-\pha{i}_2}{C_f},
 \label{eq:vcdot_impl}\\
 \dot{\pha{i}}_{2} &= \frac{K_{tr}\pha{v}_c - R_2 \pha{i}_2 - \pha{v}_{\mathrm{b}}}{L_2}.
 \label{eq:di2v_impl}
\end{align}
The proposed stand-alone voltage-control target signal is
\begin{equation}
 \pha{g}_{\mathrm{off}}= \ddot{\pha{v}}_c^{\star} - k_{v1}(\dot{\pha{v}}_c-\dot{\pha{v}}_c^{\star}) - k_{v0}(\pha{v}_c-\pha{v}_c^{\star}),
 \label{eq:gv_impl}
\end{equation}
which yields the stand-alone voltage-control law
\begin{equation}
 \pha{u}_{\mathrm{off}} = \pha{v}_c + R_1 \pha{i}_1 + L_1\dot{\pha{i}}_{2} + L_1C_f \pha{g}_v.
 \label{eq:u_off}
\end{equation}

\subsection{A new grid-connected current-control law}
The second contraction-derived SCC law is the grid-connected current-control law. It assigns the third-order target dynamics of $\pha{i}_2$ so that the inverter directly tracks the grid-service current command after reconnection. The current derivatives used in the implementation are evaluated from the measured states and the filtered inverter-side breaker-voltage estimate as
\begin{align}
 \dot{\pha{i}}_{2} &= \frac{K_{tr}\pha{v}_c - R_2 \pha{i}_2 - \pha{v}_b}{L_2},
 \label{eq:di2c_impl}\\
 \ddot{\pha{i}}_{2} &= \frac{\frac{K_{tr}}{C_f}(\pha{i}_1-\pha{i}_2)-R_2\dot{\pha{i}}_{2}-\widehat{\dot{\pha{v}}}_b}{L_2}.
 \label{eq:ddi2c_impl}
\end{align}
The proposed grid-connected current-control target signal is
\begin{equation}
 \pha{g}_{\mathrm{on}} = \widehat{\pha{i}}_2^{\star(3)} - k_{i2}(\ddot{\pha{i}}_{2}-\widehat{\ddot{\pha{i}}}_2^{\star}) - k_{i1}(\dot{\pha{i}}_{2}-\widehat{\dot{\pha{i}}}_2^{\star}) - k_{i0}(\pha{i}_2-\pha{i}_2^{\star}),
 \label{eq:gi_impl}
\end{equation}
and the corresponding grid-connected current-control law is
\begin{equation}
 \pha{u}_{\mathrm{on}} = \pha{v}_c + R_1 \pha{i}_1 + L_1\dot{\pha{i}}_{2} + \frac{L_1C_f}{K_{tr}}\left(L_2 \pha{g}_i + R_2\ddot{\pha{i}}_{2} + \widehat{\ddot{\pha{v}}}_b\right).
 \label{eq:u_on}
\end{equation}

\noindent\textit{Note:}
Since the controller is implemented in a digital average-value model, ideal differentiation is not used directly. Instead, a first-order filtered differentiator $D_{\tau}(s)=s/(\tau_d s+1)$ is used, where $\tau_d$ is selected according to the bandwidth of measurement noise. 

\subsection{Tuning of the desired tracking dynamics}
\label{subsec:tuning_desired_dynamics}

For each representative phase channel, define the tracking error as $e=x-r$. The proposed stand-alone voltage-control law is tuned first. It uses the second-order desired error dynamics
\begin{equation}
	\ddot e + k_{v1}\dot e + k_{v0}e = 0,
	\label{eq:2nd_error_dyn}
\end{equation}
where the voltage-control gains are selected from the desired bandwidth $\omega_n$ and damping ratio $\zeta$ as $k_{v1}=2\zeta\omega_n$ and $k_{v0}=\omega_n^2$.
Here, increasing $\omega_n$ speeds up the voltage recovery, while increasing $\zeta$ reduces oscillation. A critically damped choice, e.g., $\zeta=1$, is a convenient starting point for the stand-alone voltage-control law.

The proposed grid-connected current-control law is then tuned from the third-order desired error dynamics
\begin{equation}
	e^{(3)} + k_{i2}\ddot e + k_{i1}\dot e + k_{i0}e = 0 .
	\label{eq:3rd_error_dyn}
\end{equation}
To tune this law, start by choosing positive real pole rates $\lambda_1,\lambda_2,\lambda_3>0$, and then set $k_{i2}=\lambda_1+\lambda_2+\lambda_3$, $k_{i1}=\lambda_1\lambda_2+\lambda_1\lambda_3+\lambda_2\lambda_3$, and $k_{i0}=\lambda_1\lambda_2\lambda_3$. Larger $\lambda_i$ gives faster current convergence. The repeated-pole choice $\lambda_1=\lambda_2=\lambda_3$ is a good starting point because it gives a simple non-oscillatory dynamics. In principle, the pole rates can be calculated from a prescribed settling time or convergence-rate requirement, with the practical bandwidth limits caused by parameter uncertainty, sampling, derivative estimation, and measurement noise taken into account. Alternatively, one can directly use the following engineering tuning procedure: start from moderate $\omega_n$ and $\lambda_i$, then increase them until the desired recovery time is reached. If excessive derivative-filter oscillation, current ripple, or current surge is observed, the recovery time or the derivative-filter time constant $\tau_d$ should be increased.
\begin{figure}[!t]
	\centering
	\includegraphics[width=0.92\columnwidth]{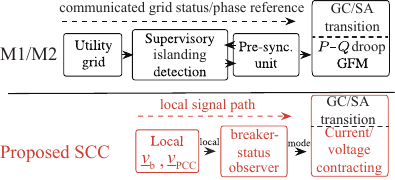}
	\caption{Comparison between the traditional communicated transition structure and the proposed breaker-status-observer-based transition structure.}
	\label{fig:baseline_droop_process}
\end{figure}
\begin{figure}[!t]
	\centering
	\includegraphics[width=0.92\columnwidth]{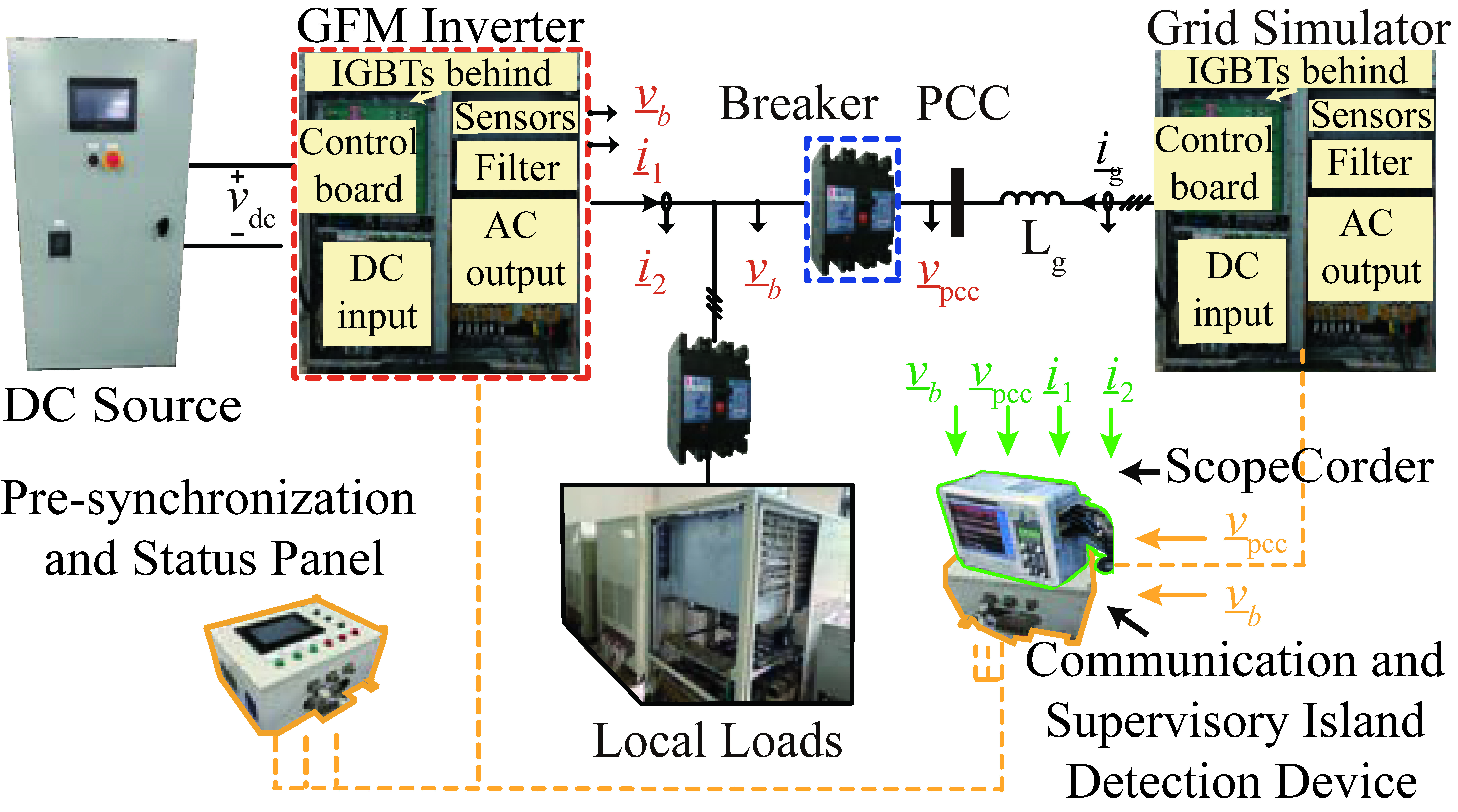}
	\caption{Laboratory-based hardware setup for the system.}
	\label{fig:hardware_prototype}
\end{figure}
\subsection{Proposed Breaker-status observer}
The breaker-status observer output $\pha{\sigma}$ selects the proposed grid-connected current-control law or stand-alone voltage-control law. 
Fig.~\ref{fig:baseline_droop_process} compares the traditional communicated transition structure with the proposed breaker-status-observer-based structure. The traditional communicated P-Q droop method relies on pre-synchronization before grid-supporting transfer, while the mode-dependent droop benchmark requires information from the islanding detector. In contrast, the proposed SCC uses the breaker-status observer to infer the electrical state from local breaker-side voltage measurements and then selects the contraction-based grid-connected current-control or stand-alone voltage-control law after the breaker event.
The breaker-status observer obtains $\pha{\sigma}$ from the local breaker two-side voltage mismatch. When the breaker is closed, the grid-side and inverter-side voltages are constrained by the same node and their mismatch remains small; after opening, the two sides are governed by different networks and the mismatch increases. To improve the robustness of the $\pha{\sigma}$ estimate, each phase mismatch is held, filtered by an $N_w$-point moving average, and accepted only after the averaged mismatch remains below $E_{\mathrm{th}}$ for $N_c$ consecutive samples. The estimator then uses the following three-phase agreement for closing: $\pha{\sigma}$ is set to $[1,1,1]^{\trans}$ only after all three phases indicate the closed condition; otherwise, it remains $[0,0,0]^{\trans}$. After the closed condition is established, each phase of $\pha{\sigma}$ is updated independently according to its own persistence-qualified voltage-mismatch result. When all three phases of $\pha{\sigma}$ return to $[0,0,0]^{\trans}$, the open condition is indicated. A short moving-average window with only a few sampling points is adopted, resulting in an observer delay of approximately $1~\mathrm{ms}$. The implemented values are listed in Table~\ref{tab:controller_param_compare}.

\begin{figure}[!t]
	\centering
	\includegraphics[width=0.82\columnwidth]{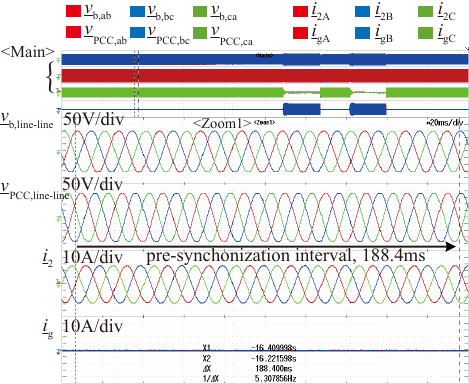}
	\caption{Measured pre-synchronization preparation interval of the M1 traditional P-Q droop method before grid-supporting transfer.}
	\label{fig:presync_record}
\end{figure}
\begin{table}[!t]
	\caption{Parameters for the Three-Method Comparison}
	\label{tab:controller_param_compare}
	\centering
	\setlength{\tabcolsep}{1.5pt}
	\renewcommand{\arraystretch}{0.98}
	\scriptsize
	\begin{tabular}{%
			p{8mm}!{\vrule width 0.5pt}%
			p{12mm}!{\vrule width 0.5pt}%
			>{\raggedright\arraybackslash}p{33mm}%
			>{\raggedleft\arraybackslash}p{18mm}%
			p{10mm}}
		\toprule
		\textbf{Group} & \textbf{Symbol} & \textbf{Description} & \textbf{Value} & \textbf{Units} \\
		\midrule
		\multirow{10}{*}{\rotatebox{90}{\textbf{System}}}
		& $S$ & Rated apparent power of inv. & 5 & kVA \\
		& $L_1$ & Inv.-side inductance & 0.3 & mH \\
		& $C_f$ & Filter capacitance & 345 & $\mu$F \\
		& $L_2$ & Grid-side filter inductance & 2.5 & mH \\
		& $K_{tr}$ & Transformer ratio & 1.046 & -- \\
		& $V_{\mathrm{PCC}}$ & PCC rated line-line voltage rms & 120 & V \\
		& $f_g$ & Nominal grid freq. & 50 & Hz \\
		& $f_{\mathrm{sw}}, f_s$ & Switching freq., Sampling freq. & 7.8 & kHz \\
		&$L_g$ &  Grid-side line inductance & 0.4 & mH \\
		& $P_{\mathrm{L}},Q_{\mathrm{L}}$ & Full local load & 3, 3 & kW, kVAr \\
		\cmidrule(lr){1-5}
		\multirow{8}{*}{\rotatebox{90}{\textbf{M1 P-Q droop~\cite{vandoorn2013transition_droop}}}}
		& $P_{\mathrm{ref}}$ & Active-power ref. & 1.5 & kW \\
		& $Q_{\mathrm{ref}}$ & Reactive-power ref. & 1.5 & kVAr \\
		& $V_{\mathrm{nom}}$ & Nominal voltage & 120 & V \\
		& $f_{\mathrm{nom}}$ & Nominal frequency & 50 & Hz \\
		& $K_Q$ & $Q$--$V$ droop gain & $2.0\times10^{-4}$ & V/VAr \\
		& $K_P$ & $P$--$f$ droop gain & 0.128 & rad/(s\,W) \\
		& $K_{p,u}$ & Voltage-loop P gain & $1.875\times10^{-5}$ & V$^{-1}$ \\
		& $K_{i,u}$ & Voltage-loop I gain & $3.5287\times10^{-5}$ & V$^{-1}$s$^{-1}$ \\
		\cmidrule(lr){1-5}
		\multirow{10}{*}{\rotatebox{90}{\textbf{M2 mode-dep.~\cite{ref7}}}}
		& $P_{\mathrm{ref}}$ & Effective Active-power ref. & 1.5  & kW \\
		& $Q_{\mathrm{ref}}$ & Effective Reactive-power ref. & 1.5 & kVAr \\
		& $V_{\mathrm{nom}}$ & Nominal voltage & 120 & V \\
		& $f_{\mathrm{nom}}$ & Nominal frequency & 50 & Hz \\
		& $n$ & $P$ droop gain & 0.6 & \tabincell{c}{rad/s/kW} \\
		& $m$ & $Q$ droop gain & 40 & \tabincell{c}{V/kVAr} \\
		& $m_{\mathrm{int}}$ & $Q$ integral gain & 0.02 & \tabincell{c}{V/s/kVAr} \\
		& $\tau_p$ & Power-filter time & 33 & ms \\
		& $K_{p,u}$ & Voltage-loop P gain & $1.875\times10^{-5}$ & V$^{-1}$ \\
		& $K_{i,u}$ & Voltage-loop I gain & $3.5287\times10^{-5}$ & V$^{-1}$s$^{-1}$ \\
		\cmidrule(lr){1-5}
		\multirow{11}{*}{\rotatebox{90}{\textbf{Proposed SCC}}}
		& $\hat L_1$ & Est. inv.-side inductance & 0.25 & mH \\
		& $\hat L_2$ & Est. grid-side inductance & 2.1 & mH \\
		& $\hat C_f$ & Est. filter capacitance & 345 & $\mu$F \\
		& $\tau_d$ & Derivative-filter delay & 384.615 & $\mu$s \\
		& $E_{\mathrm{th}}$ & Breaker-status threshold & 6.0 & V \\
		& $N_w,N_c$ & Observer window/count & 3, 10 & samples \\
		& $\lambda_{1,2,3}$ & GC current-law pole rate & $2.03\times10^3$ & s$^{-1}$ \\
		& $\omega_n$ & Voltage-loop bandwidth & 2800.617 & s$^{-1}$ \\
		& $\zeta$ & Voltage-loop damping ratio & 0.5 & -- \\
		& $k_{v1}$ & $\dot{v}_c$ error gain & $1.3882\times10^3$ & s$^{-1}$ \\
		& $k_{v0}$ & $v_c$ error gain & $5.3528\times10^6$ & s$^{-2}$ \\
		\bottomrule
	\end{tabular}
	\normalsize
\end{table}
\section{Hardware Experiment and Results}
\subsection{Hardware configuration and test conditions}
The hardware test system in Fig.~\ref{fig:hardware_prototype} uses the same inverter power stage, $L_1C_fL_2$ filter, breaker, controller platform, sensing chain, grid interface, and local load for all compared methods. \textbf{M1} is the traditional communicated P--Q droop implementation shown in Fig.~\ref{fig:baseline_droop_process}, where the stand-alone-to-grid-connected transition is performed through a communicated pre-synchronization process shown in Fig.~\ref{fig:presync_record}, a $188.4~\mathrm{ms}$ preparation interval. \textbf{M2} is the mode-dependent droop benchmark in \cite{ref7}, implemented here with an islanding-detection device that permits reconnection only after the voltage difference satisfies the acceptable reconnection condition. The \textbf{proposed SCC} relies only on local voltage measurements. It does not require breaker-status communication or an islanding-detection process. The system parameters and controller settings are summarized in Table~\ref{tab:controller_param_compare}. 
The measured waveform channels follow the system configuration: inverter-side line-line voltages $v_{b,ab}$, $v_{b,bc}$, $v_{b,ca}$; PCC-side line-line voltages $v_{\mathrm{PCC},ab}$, $v_{\mathrm{PCC},bc}$, $v_{\mathrm{PCC},ca}$; inverter current $i_2$ positive toward the load; and grid current $i_g$ positive from the grid toward the load-side node. 
\begin{comment}
Thus, Figs.~\ref{fig:exp_case_rated}--\ref{fig:exp_case_sag} are read from top to bottom as $v_b$, $v_{\mathrm{PCC}}$, $i_2$, and $i_g$, with red/blue/green traces denoting $AB/BC/CA$ voltage channels or A/B/C current phases. 
\end{comment}

\begin{figure*}[!t]
	\centering
	\includegraphics[width=0.95\textwidth]{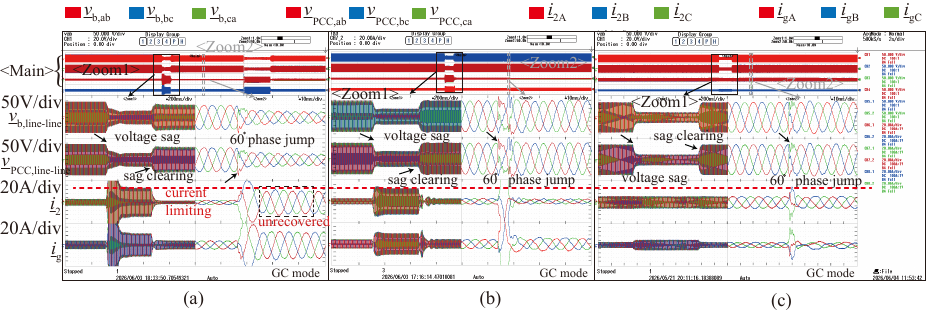}
	\caption{Experimental results of grid-connected-mode comparison under large grid-voltage disturbances, including voltage sag and $60^\circ$ phase jump: (a) M1 P-Q droop, current limiting after sag, large fluctuation, current limiting after phase jump; (b) M2 mode-dependent droop, current limiting after sag, large current transient but recovered after phase jump; (c) proposed SCC, stable controlled current with no current limiting under sag, contracting current recovery after phase jump, lower current fluctuation.}
	\label{fig:exp_case_sag}
\end{figure*}
\begin{figure*}[!t]
	\centering
	\includegraphics[width=0.95\textwidth]{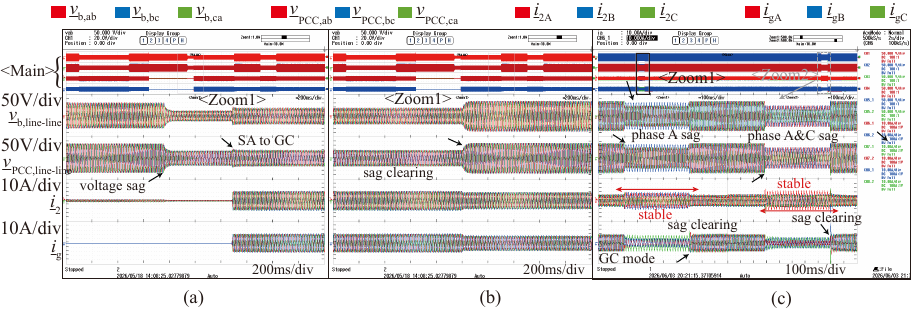}
	\caption{Experimental results of the proposed SCC under large-signal disturbances: (a) stand-alone $\rightarrow$ grid-connected transition for grid support under sag fault, $0.31+\mathrm{j}0.01\rightarrow1.38+\mathrm{j}1.06$~kVA, demonstrating grid-support power under sag; (b) sag clearing during grid-connected operation, grid-support power injection, $1.38+\mathrm{j}1.06\rightarrow2.38+\mathrm{j}1.77$~kVA; (c) stable response under unsymmetrical fault, phase-A sag $\rightarrow$ sag clearing $\rightarrow$ phase-A/phase-C sag $\rightarrow$ sag clearing.}
	\label{fig:exp_case_sagunsymmetrical}
\end{figure*}

\begin{figure*}[!t]
	\centering
	\includegraphics[width=0.95\textwidth]{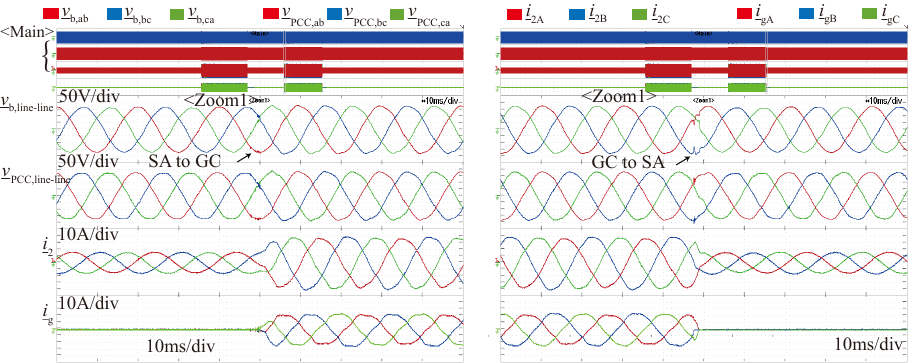}
	\caption{Experimental results of the proposed SCC under unplanned seamless overload transition with large operating-point change: local-load support and reactive-power injection to the grid in grid-connected mode, bidirectional complex-power transition $1.6+\mathrm{j}1.6\leftrightarrow2.2+\mathrm{j}5.4$~kVA, current magnitude about $0.45\rightarrow1.17$~p.u. ($260\%$ change), demonstrating transient stability.}
	\label{fig:exp_case_half_full}
\end{figure*}

\begin{figure*}[!t]
	\centering
	\includegraphics[width=0.95\textwidth]{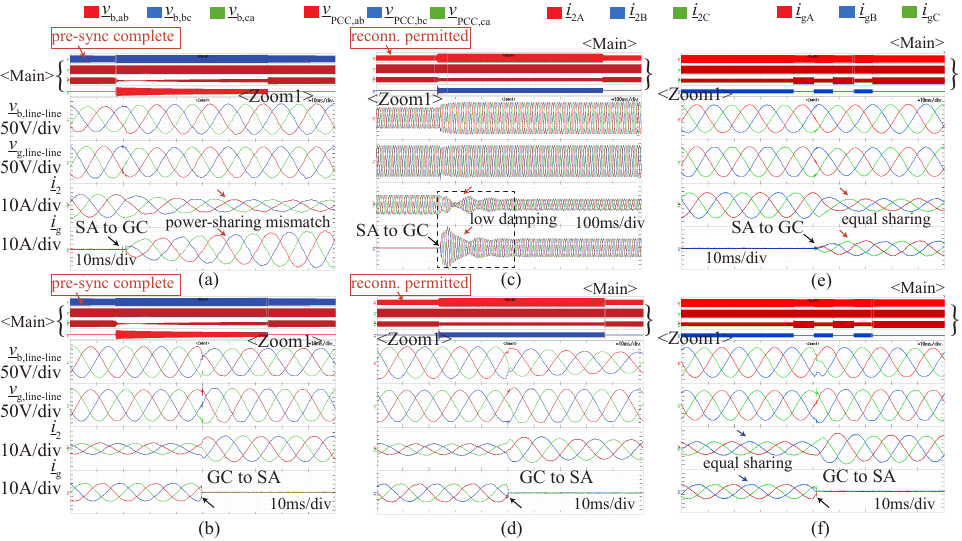}
	\caption{Experimental results of full local-load inverter-grid power-sharing comparison. (a) M1 P-Q droop, just after SA-to-GC, $P=1.10$~kW, $Q=0.80$~kVAr; (b) M1 just before GC-to-SA, $P=1.52$~kW, $Q=1.10$~kVAr; (c) M2 mode-dependent droop, just after SA-to-GC, $P=1.60$~kW, $Q=1.20$~kVAr; (d) M2 just before GC-to-SA, $P=1.57$~kW, $Q=1.24$~kVAr; (e) proposed SCC without reconnection preparation, just after SA-to-GC, $P=1.67$~kW, $Q=1.51$~kVAr; (f) proposed SCC just before GC-to-SA, $P=1.68$~kW, $Q=1.50$~kVAr, showing fast sharing recovery.}
	\label{fig:exp_case_rated}
\end{figure*}

\subsection{Experimental validation waveforms}
The hardware waveforms are presented in the order of grid-voltage-disturbance comparison, proposed-SCC large-signal disturbance results, overload grid support, and full-local-load sharing. Fig.~\ref{fig:exp_case_sag} compares the three methods in grid-connected mode under the same voltage sag and $60^\circ$ phase-jump disturbance: Fig.~\ref{fig:exp_case_sag}(a) for M1 P-Q droop, Fig.~\ref{fig:exp_case_sag}(b) for M2 mode-dependent droop, and Fig.~\ref{fig:exp_case_sag}(c) for the proposed SCC. For M1, the sag triggers current limiting, and the phase jump causes large unstable current fluctuation with current limiting again triggered. For M2, the sag also triggers current limiting, while the phase jump produces a large current transient that recovers without entering current limiting after the jump. In contrast, the proposed SCC maintains stable controlled current without current limiting during the sag, and after the phase jump the current contracts back to the reference within about 5~ms with lower fluctuation.

Fig.~\ref{fig:exp_case_sagunsymmetrical} presents proposed-SCC results for grid support under sag and stable operation under unsymmetrical faults. Fig.~\ref{fig:exp_case_sagunsymmetrical}(a) shows stand-alone-to-grid-connected transition for grid support under sag fault, with the measured complex power changing from $0.31+\mathrm{j}0.01$ to $1.38+\mathrm{j}1.06$~kVA. Fig.~\ref{fig:exp_case_sagunsymmetrical}(b) shows sag clearing during grid-connected operation and grid-support power injection, with the measured complex power changing from $1.38+\mathrm{j}1.06$ to $2.38+\mathrm{j}1.77$~kVA. The complex-power increases in Figs.~\ref{fig:exp_case_sagunsymmetrical}(a) and (b) show that the inverter can actively support the grid under the sag condition. Fig.~\ref{fig:exp_case_sagunsymmetrical}(c) shows stable operation even under an unsymmetrical fault sequence: phase-A sag, sag clearing, phase-A/phase-C sag, and sag clearing.

Fig.~\ref{fig:exp_case_half_full} verifies overload transition capability under the proposed SCC with a large operating-point change. In grid-connected mode, the inverter supports the local load and injects reactive power to the grid. The approximate bidirectional power transition is $1.6+\mathrm{j}1.6\leftrightarrow2.2+\mathrm{j}5.4$~kVA, corresponding to a current-magnitude change from about $0.45$ to $1.17$~p.u. ($260\%$ change). This test demonstrates transient stability under a large post-transition operating-point shift and highlights that the current-control law can track a large current-vector change during the mode transition; the stand-alone-to-grid-connected overload transition is the more demanding direction because the inverter must quickly increase grid-support current after breaker closing.

Fig.~\ref{fig:exp_case_rated} verifies the full-local-load condition, where the inverter and grid should share the local load in GC operation and the inverter should support the full load in SA operation. Just after SA-to-GC transfer and just before GC-to-SA transfer, the measured GC inverter-side powers $P+\mathrm{j}Q$ are $1.10+\mathrm{j}0.80$ and $1.52+\mathrm{j}1.10$~kVA for M1, $1.60+\mathrm{j}1.20$ and $1.57+\mathrm{j}1.24$~kVA for M2, and $1.67+\mathrm{j}1.51$ and $1.68+\mathrm{j}1.50$~kVA for the proposed SCC. These GC powers show that the proposed SCC reaches the equal power-sharing condition rapidly and accurately after the unplanned transitions.

\begin{table}[!t]
	\caption{Representative Seamless-Transfer Studies and Proposed SCC}
	\label{tab:lit_compare}
	\centering
	\scriptsize
	\setlength{\tabcolsep}{2pt}
	\renewcommand{\arraystretch}{1.08}
	\begin{tabularx}{\columnwidth}{@{}>{\raggedright\arraybackslash}p{0.28\columnwidth}>{\centering\arraybackslash}p{0.12\columnwidth}>{\raggedright\arraybackslash}p{0.28\columnwidth}>{\raggedright\arraybackslash}X@{}}
		\toprule
		Method & Direction & Transition Method & Feature \\
		\midrule
		Droop-based resynchronization \cite{vandoorn2013transition_droop,amin2020resynchronization_udc,deng2017enhanced_power_flow} & SA$\rightarrow$GC & Voltage/frequency PS interval & Planned return \\
		Droop synchronization for parallel inverters \cite{ramezani2020seamless} & SA$\rightarrow$GC & Torque/droop PS interval & Planned return \\
		LQR/supplementary transition control \cite{ganjian2021lqr_transition,azimi2021supplementary_transition} & Bi. & Transfer command; PS interval & Commanded transfer; no GC-mode large-disturbance validation \\
		Critical-infrastructure droop \cite{ref7} & Bi. & Islanding detection needed; no PS interval & Reconnection waiting required \\
		Pre-synchronization-unit-aided hybrid/droop control \cite{ref10,arafat2015smooth_transition} & Bi. & Network unit; PS interval & Requires PS interval; long recovery \\
		Unified control scheme \cite{ref8} & GC$\rightarrow$SA & Local mode logic; no PS interval & Islanding only \\
		Microgrid transition method \cite{ref9} & GC$\rightarrow$SA & Islanding detection; no PS interval & Islanding only; no GC-mode large-disturbance validation \\
		Phase-adaptive current limiting \cite{guo2026seamless_gfm_pv} & SA$\rightarrow$GC & Current limit; no PS interval & Current-limited reconnection; current fluct. \\
		Proposed SCC & Bi. & Local voltage; no PS interval & Unplanned; contracting dynamics; stable under large disturbances \\
		\bottomrule
	\end{tabularx}
	\normalsize
\end{table}

The proposed SCC response is faster because the proposed grid-connected current-control law tracks injected current after the breaker-status observer identifies the electrical state. The assigned contraction rates give design settling scales of $4/\lambda=1.97~\mathrm{ms}$ for the current-control law and $4/(\zeta\omega_n)=5.76~\mathrm{ms}$ for the stand-alone voltage-control law. Across Figs.~\ref{fig:exp_case_sag}, \ref{fig:exp_case_sagunsymmetrical}, \ref{fig:exp_case_half_full}, and \ref{fig:exp_case_rated}, the measured proposed-SCC responses all satisfy these assigned settling-scale conditions, with current envelopes settling in about 4--5~ms and inverter-side voltages returning to steady sinusoidal envelopes in about 2~ms.

The droop-based baselines require either transition preparation or droop-loop settling before reaching the post-event power target. M1 follows the communicated pre-synchronization path shown in Fig.~\ref{fig:baseline_droop_process}; the recorded waveform in Fig.~\ref{fig:presync_record} shows a $188.4~\mathrm{ms}$ preparation interval before grid-supporting transfer. M2 is governed by the droop power loop and requires islanding detection before enabling breaker closing. In both cases, the system must recover either from a phase-aligned pre-synchronization condition or from near-zero power support to the corresponding post-event equilibrium, which delays power support during unplanned transitions. Overall, the experiments verify that the proposed SCC with the breaker-status observer provides local-load voltage support after breaker opening and rapid grid-service power injection after reconnection, without requiring breaker-status communication or a dedicated pre-synchronization interval. Table~\ref{tab:lit_compare} summarizes the comparison between representative studies reviewed in the Introduction and the proposed SCC.

\section{Conclusion}

In this article, a seamless contraction-control (SCC) framework for target dynamics is proposed for stable large-signal voltage/current behavior of an $LCL$-filtered grid-forming inverter during unplanned grid-connected/stand-alone transitions. Based on the SCC, contraction-based current-control and voltage-control laws are derived, together with controller-parameter tuning methods for assigning the desired convergence dynamics. A breaker-status observer is also integrated with the SCC so that the inverter can realize unplanned bidirectional grid-connected/stand-alone transitions without breaker-status communication, supervisory islanding detection, or a dedicated pre-synchronization interval. Experimental comparisons under symmetrical grid-voltage sag and phase jump show that the proposed SCC provides smoother voltage/current responses than the communicated P--Q droop and mode-dependent droop controllers. Experimental comparisons of the full-local-load sharing case show that, after grid connection, the proposed SCC reaches the desired inverter-grid power-sharing condition faster and with the assigned convergence rate. Additional results for unsymmetrical grid-voltage sag, unplanned SA-to-GC/GC-to-SA transitions during symmetrical voltage sag, and overload transition with a large operating-point change show stable operation and fast SCC convergence rate. Compared with communicated P--Q droop and mode-dependent droop, the proposed SCC reduces the dependence on transition preparation and improves recovery behavior for unplanned bidirectional transitions under large-signal disturbances.

\bibliographystyle{IEEEtran}
\bibliography{IEEEabrv,Bibliography}

\end{document}